\documentclass{JHEP3} % 10pt is ignored!

\usepackage{epsfig,multicol}
\usepackage{amsmath, amssymb}
\usepackage{epic}
\usepackage{concmath, palatino}

\newcommand\fverb{\setbox\pippobox=\hbox\bgroup\verb}
\newcommand\fverbdo{\egroup\medskip\noindent%
			\fbox{\unhbox\pippobox}\ }
\newcommand\fverbit{\egroup\item[\fbox{\unhbox\pippobox}]}
\newbox\pippobox
\newcommand{\be}{\begin{equation}}
\newcommand{\ee}{\end{equation}}
\newcommand{\ba}{\begin{eqnarray}}
\newcommand{\ea}{\end{eqnarray}}

\newcommand{\refeq}[1]{Eq.~(\ref{eq:#1})}

\newcommand{\ads}{AdS_5\times S^5}
\newcommand{\tran}{transcendentality\ }

\newtheorem{theorem}{Theorem}[section]

\title{QCD properties of twist operators in the \\ ${\cal N}=6$ Chern-Simons theory}

\author{Matteo Beccaria and Guido Macorini\\
  Physics Department, Salento University, 
  Via Arnesano, 73100 Lecce\\
  INFN, Sezione di Lecce\\
  E-mail: \email{matteo.beccaria@le.infn.it} 
}

\abstract{
We consider twist-1, 2 operators in planar ${\cal N}=6$ superconformal Chern-Simons ABJM theory. We derive higher order anomalous dimensions
from integrability and test various QCD-inspired predictions known to hold in ${\cal N}=4$ SYM. In particular, we show that the asymptotic anomalous dimensions
display intriguing remnants of Gribov-Lipatov reciprocity and Low-Burnett-Kroll logarithmic cancellations. Wrapping effects are also discussed and shown to be 
subleading at large spin.
}

\begin{document} 

\section{Introduction}

Integrability in AdS/CFT correspondence~\cite{Maldacena:1997re} is a deep and intriguing feature linking the gauge and string theory dynamics.
As a welcome tool, it offers the worth possibility of performing unexpectedly easy higher order calculations on the gauge theory side. 
This major outcome opens the way to the investigation of hidden properties of the perturbative expansion. An important example
of these general considerations is ${\cal N}=4$ SYM in $AdS_5/CFT_4$ duality, at least in the planar 't Hooft limit.
The pure gauge sector is universal and independent on the large amount of supersymmetry. Thus, it is sensible to test QCD-inspired physical conjectures
which find their origin in the gauge dynamics. In particular, one can 
consider the so-called twist operators belonging to the $\mathfrak{sl}(2)$ sector of ${\cal N}=4$ SYM. These are single trace composite operators with 
arbitrarily large Lorentz spin $N$ carried by covariant derivatives. In QCD, twist-2 anomalous dimensions $\gamma(N)$ enter the evolution equations of deep inelastic scattering (DIS)
and are physically very relevant~\cite{Altarelli:1981ax}.
The analysis of $\gamma(N)$ suggests, among others, the following three basic predictions about the large $N$ (quasi-elastic) regime:
\begin{enumerate}
\item[(a)] {\em Logarithmic scaling.} The large $N$ dominant term is logarithmic, $\gamma(N) \sim f(g)\,\log N$, where $f(g)$ is the universal cusp anomalous 
dimension~\cite{Korchemsky:1992xv,Belitsky:2003ys,Belitsky:2006en,Alday:2007mf}.

\item[(b)] {\em Gribov-Lipatov reciprocity.} This is a crossing relation which implies an infinite set of constraints on the subleading terms 
appearing in the large $N$ expansion of $\gamma(N)$~\cite{Dokshitzer:2005bf,Dokshitzer:2006nm,Basso:2006nk}.
% The function $P(N)$ defined implicitly by $\gamma(N) = P(N+\frac{1}{2}\,\gamma(N))$ can be expanded at large $N$
% and involves only inverse integer powers of the collinear Casimir $J^2 = N(N+1)$ with logarithmic enhancement factors.

\item[(c)] {\em Low-Burnett-Kroll wisdom.} The anomalous dimension $\gamma(N)$ develops high powers of $\log N$ increasing with the perturbative order, the leading terms
being of the form $(\log(N)/N)^k$. Nevertheless, many of these terms are {\em inherited} from lower order calculations. This is independent on (b)
and can be traced back to quite general old results simply related to gauge invariance~\cite{Low:1958sn}.
%
%, many cancellations are expected 
%t the level of $P(N)$. In turns, this means that there are many hidden relations between the logarithms appearing in $\gamma(N)$.
\end{enumerate}

The status of these predictions can be looked upon in the perspective of AdS/CFT duality. 
Logarithmic scaling, property (a), is well established and understood both in gauge theory~\cite{Belitsky:2006en} and in string theory~\cite{Frolov:2002av}.
Quantitatively, the function $f(g)$ is predicted at all orders in the weak coupling expansion~\cite{Beisert:2006ez} as well as at strong coupling~\cite{Basso:2007wd}.
Gribov-Lipatov reciprocity, property (b), can be understood in terms of crossing symmetry between DIS and its time-like counter part, {\em i.e.}
$e^+e^-$ annihilation~\cite{Dokshitzer:2005bf}.
It is well tested in the gauge theory up to 5 loops and, surprisingly, it admits wide generalizations
extending in some form to larger parts of 
$\mathfrak{psu}(2,2|4)$ \cite{Beccaria:2009eq,Beccaria:2009vt,Beccaria:2008fi,Beccaria:2007pb,Beccaria:2007bb,Beccaria:2007vh,Beccaria:2007cn}.
Its test in string theory is definitely non trivial as discussed in~\cite{Beccaria:2008tg}. 
The last prediction, property (c), takes its name after Low-Burnett-Kroll (LBK) theorems~\cite{Low:1958sn}. As we mentioned, 
these are old general results which improve the eikonal leading order factorization. They rely on gauge invariance alone and are of a quite general validity. As for (b)
it is difficult to test them at strong coupling lacking a systematic way to treat the (quantized) large spin limit.

Despite this admittedly involved and open picture, we immediately notice the following hierarchy among the three predictions.
Property (a) has a very simple origin both in the gauge theory where it is related to the universal form of soft gluon emission and
in string theory where it is linked to the large spin regime of rotating semiclassical strings. On the opposite side, 
property (b) is very specific to the SYM gauge dynamics and is rather unclear on the string side. 
Finally, property (c) stands somewhat in the middle.

Inspired by~\cite{Agarwal:2008pu}, we believe that these are valid motivations for investigating the QCD-inspired properties (a), (b) and (c) moving from ${\cal N}=4$ SYM
to the ABJM theory~\cite{Aharony:2008ug}. This is a three dimensional $U(N)\times U(N)$ gauge theory with four complex scalars in the $(N, \overline N)$ representation, 
their fermionic partners, and a Chern-Simons action with levels $+k$, $-k$. This theory has ${\cal N}=6$ superconformal symmetry $\mathfrak{osp}(2,2|6)$.
ABJM can be considered as the low energy theory of $N$ parallel M2-branes at a $\mathbb{C}^4/\mathbb{Z}_k$ singularity. In the large $N$ 
limit this is M theory on $AdS_4\times S^7/\mathbb{Z}_k$. For fixed $\lambda=N/k$ we can describe it by type IIA string on $AdS_4\times \mathbb{CP}^3$
which is classically integrable~\cite{Arutyunov:2008if,Stefanski:2008ik,Gomis:2008jt}.
The manifest (non abelian) part of the R symmetry is $SU(2)\times SU(2)$. The complex scalars can be written as two  doublets transforming as $(2,1)$ and $(1,2)$.
Under the gauge group they transform as $(N, \overline N)$ and $(\overline N, N)$.
At leading order (two loops, $\lambda^2$ in 't Hooft coupling $\lambda$), the dilatation operator for single trace operators built with 
these scalars is a $SU(4)$ integrable spin chain~\cite{Minahan:2008hf,Bak:2008cp}.
In~\cite{Gromov:2008qe}, Gromov and Vieira have proposed a set of all-loop Bethe-Ansatz equations for the full $\mathfrak{osp}(2,2|6)$ theory consistent with the 2 loop analysis and with the 
superstring algebraic curve at strong coupling~\cite{Gromov:2008bz}. The equations depend on a dressed coupling $h(\lambda)$ which takes into account the fact that 
the one-magnon dispersion relation is not protected~\cite{Nishioka:2008gz,Gaiotto:2008cg,Grignani:2008is,McLoughlin:2008he}.
Finally,  an exact  $SU(2|2)_A \oplus SU(2|2)_B$ symmetric $S$-matrix consistent with~\cite{Gromov:2008qe} has been presented in~\cite{Ahn:2008aa}.

The important point for our investigation is that twist operators can be found in a $\mathfrak{sl}(2)$-like sector of ABJM as discussed in~\cite{Gromov:2008qe,Zwiebel:2009vb}.
At strong coupling and large spin they behave quite similarly to the corresponding ones in $\ads$. In particular their dual string state is a folded string 
rotating in $AdS_3$ with large spin $N$ and with angular momentum $J\sim \log\,N$ in $\mathbb{CP}^3$~\cite{folded3} in close analogy to the well known 
folded string solution in $\ads$~\cite{Frolov:2002av}. At weak coupling, they are composite operators in totally different theories. Nevertheless, 
both ${\cal N}=6$ SCS and ${\cal N}=4$ SYM are integrable and the all-loop Bethe equations in the $\mathfrak{sl}(2)$ sectors are very close.
Also, from the leading order analysis of twist-1 operators~\cite{Zwiebel:2009vb}, it seems that maximal transcendentality Ans\"atze are feasible.
In conclusion, it is very natural to attempt to determine multi-loop anomalous dimensions of twist operators in ABJM and see whether the structural similarity 
with ${\cal N}=4$ is powerful enough to preserve some remnant of the (a)-(b)-(c) properties. 

As is well known, in perturbation theory, the spin $N$ dependent anomalous dimension $\gamma(N)$ is the sum of two pieces, the {\em asymptotic} and {\em wrapping} contributions.
The asymptotic term can be computed rigorously for each $N$ by means of the  all loop Bethe Ansatz equations~\cite{Gromov:2008qe}.
The wrapping correction  starts at a twist dependent loop order and takes into account finite volume corrections to multi-particle states.
In ${\cal N}=4$, the above QCD-inspired properties holds separately for the asymptotic and wrapping contributions.

In this paper, we shall derive several higher order closed expressions for the asymptotic part of the anomalous dimensions of ABJM 
twist-1 and twist-2 operators. We shall derive these results by 
combining analytical results based on a suitable Baxter equation and the maximal transcendentality Ansatz so fruitful in the ${\cal N}=4$ case.
The obtained expressions can be analyzed in the spirit of looking for special reciprocity or LBK cancellation features.
Concerning wrapping effects, they are currently believed to be correctly predicted in ${\cal N}=4$ by generalized 
L\"{u}scher formulas~\cite{Janikprev}. For the ABJM theory, we shall present some results obtained in the framework 
of the recent proposal~\cite{Gromov:2009tv} (see also the developments~\cite{Bombardelli:2009ns,Arutyunov:2009ur}).

\section{Twist operators in the $\mathfrak{sl}(2)$ sector of ABJM}

The all-loop Bethe equations for ABJM has been proposed in~\cite{Gromov:2008qe}. They can be concisely and conveniently summarized by the 
following $\mathfrak{osp}(2,2|6)$ Dynkin diagram (associated with the fermionic $\eta=-1$ grading)
\be
\begin{minipage}{260pt}
\setlength{\unitlength}{1pt}
\small\thicklines
\begin{picture}(160,80)(-60,-50)
%
%\dottedline{3}(-32,0)(-10,0)  % initial dotted line 
%\put(-32,0){\line(1,0){22}}  % initial solid line 
\put(  0,00){\circle{15}}
\put( -5,-5){\line(1, 1){10}}  % cross on node 1
\put( -5, 5){\line(1,-1){10}}  % cross on node 1
%\put(  0,15){\makebox(0,0)[b]{$K_1$}}  % weight on node 1
\put(  0,-15){\makebox(0,0)[t]{$u_1$}} % roots on node 1
\put(  7,00){\line(1,0){26}} % 1-2 solid
%\dottedline{3}(8,0)(32,0)    % 1-2 dotted
\put( 40,00){\circle{15}}     
%\put( 35,-5){\line(1, 1){10}}  % cross on node 2
%\put( 35, 5){\line(1,-1){10}}  % cross on node 2
%\put( 40,15){\makebox(0,0)[b]{$K_2$}} % weight on node 2
\put( 40,-15){\makebox(0,0)[t]{$u_2$}} % roots on node 2
\put( 47,00){\line(1,0){26}} % 2-3 solid
%\dottedline{3}(48,0)(72,0)   % 2-3 dotted 
\put( 80,00){\circle{15}}
\put( 75,-5){\line(1, 1){10}}  % cross on node 3
\put( 75, 5){\line(1,-1){10}}  % cross on node 3
%\put( 80,15){\makebox(0,0)[b]{$K_3$}} % weight on node 3
\put( 80,-15){\makebox(0,0)[t]{$u_3$}}  % roots on node 3
\put( 88,00){\line(3,2){25}} % 3-4 solid
\put(120,20){\circle{15}}
\put(115,15){\line(1, 1){10}}  % cross on node 3
\put(115,25){\line(1,-1){10}}  % cross on node 3
\put( 88,00){\line(3,-2){25}} % 3-4 solid
\put(120,-20){\circle{15}}
\put(115,-25){\line(1, 1){10}}  % cross on node 3
\put(115,-15){\line(1,-1){10}}  % cross on node 3
\put(150,18){\makebox(0,0)[b]{$u_4$}}
\put(150,-28){\makebox(0,0)[b]{$u_{\overline 4}$}}
\put(118,13){\line(0,-1){25}} % 3-4 solid
\put(122,13){\line(0,-1){25}} % 3-4 solid
\end{picture}
\end{minipage}
\ee
We shall consider twist operators in the $\mathfrak{sl}(2)$ sector where we excite symmetrically 
the same number $N$ of $u_4$ and $u_{\overline{4}}$ roots. An explicit description of these states as single trace 
composite operators with length $2L$ can be found in~\cite{Zwiebel:2009vb}. As in the ${\cal N}=4$ case, 
the integer $L$ can be identified with the twist of the operator.

The Bethe equations involve the deformed spectral parameters $x^\pm$ defined by 
\be
x^\pm + \frac{1}{x^\pm} = \frac{1}{h}\left(u\pm \frac{i}{2}\right),
\ee
where $h(\lambda)$ is the interpolating coupling appearing in the one-magnon dispersion relation. 
For twist $L$ operators  they read (see App.~(A.1) of ~\cite{Gromov:2008qe})
\be
\left(\frac{x^+_k}{x^-_k}\right)^L = -\prod_{j\neq k}^N\frac{u_k-u_j+i}{u_k-u_j-i}\,\left(
\frac{x_k^--x_j^+}{x_k^+-x_j^-}\right)^2\,\sigma^2_{\rm BES}.
\ee
The only difference compared with ${\cal N}=4$ SYM is the extra minus sign. This will turn out to be definitely 
relevant to our analysis~\footnote{The effect of this sign in the thermodynamical limit and at strong coupling is discussed in~\cite{McLoughlin:2008he}.}.
The factor $\sigma_{\rm BES}$ is the dressing phase which will play no role at the perturbative order explored in this paper.
A more convenient form for weak coupling expansions is 
\be
\left(\frac{x^+_k}{x^-_k}\right)^L = -\prod_{j\neq k}^N\frac{x^-_k-x^+_j}{x^+_k-x^-_j}\,
\frac{\displaystyle 1-\frac{1}{x^+_k x^-_j}}{\displaystyle 1-\frac{1}{x^-_k x^+_j}}\,\sigma^2_{\rm BES}.
\ee
Everything can be written in terms of the momentum $p(u)$ defined by
\be
p(u) = -i\,\log\frac{x^+(u)}{x^-(u)}, 
\ee
and using the useful relations 
\ba
u(p) &=& \frac{1}{2}\,\cot\frac{p}{2}\,\sqrt{1+16\,h^2\,\sin^2\frac{p}{2}}. \\
x^\pm(u(p)) &=& \frac{1}{h}\,\frac{e^{\pm\,i\,\frac{p}{2}}}{4\sin\frac{p}{2}}\left(1+\sqrt{1+16\,h^2\,\sin^2\frac{p}{2}}\right).
\ea
The contribution to the energy/anomalous dimension from the Asymptotic Bethe Ansatz (ABA) equations is 
\be
\gamma^{\rm ABA}(h) = \sum_{k=1}^N\left[\sqrt{1+16\,h^2\,\sin^2\frac{p_k}{2}}-1\right] = \sum_{n=1}^\infty \gamma_{2\,n}\,h^{2\,n}.
\ee
At large $N$, we expect $\gamma(h) \sim f_{CS}(h)\,\log\,N + \cdots$ with 
the six-loop cusp anomaly 
\be
\label{eq:cusp}
f_{CS}(h) = 4\,h^2-\frac{4}{3}\,\pi^2\,h^4+\frac{44}{45}\,\pi^4\,h^6 + \cdots.
\ee

\section{Twist-1}
\subsection{The two-loop problem}

The two-loop anomalous dimension is presented in~\cite{Zwiebel:2009vb}. Here, we derive 
this result in a different way by using the very efficient Baxter function formalism.
The Bethe Ansatz equations are 
\be
\frac{u_k+\frac{i}{2}}{u_k-\frac{i}{2}} = -\prod_{j\neq k}^N\frac{u_k-u_j-i}{u_k-u_j+i},\qquad k,j = 1, \dots, N.
\ee
The Baxter polynomial associated to the ground state is 
\be
Q_N(u) = {\cal N}\,\prod_{k=1}^N (u-u_k).
\ee
It obeys the equivalent leading order Baxter equation
\be
\left(u+\frac{i}{2}\right)\,Q_N(u+i)-\left(u-\frac{i}{2}\right)\,Q_N(u-i) = i\,(2\,N+1)\,Q_N(u).
\ee
The solution to this recurrence obeying the polynomiality condition is 
\be
Q_N(u) = {}_2F_1\left(\left.\begin{array}{c}
-N, \quad   i\,u+\frac{1}{2} \\
1
\end{array}
\right| 2 \right).
\ee
It follows that the two-loops anomalous dimension can be computed exactly and reads~\footnote{As usual, (nested) harmonic sums are recursively defined by 
$$
S_a(N) = \sum_{n=1}^N \frac{(\mbox{sign} a)^n}{n^{|a|}},\quad S_{a, b, \dots}(N) = \sum_{n=1}^N \frac{(\mbox{sign} a)^n}{n^{|a|}}\,S_{b, \dots}(n).
$$
}
\be
\label{eq:h2}
\gamma_2^{\rm ABA}(N) = \sum_k\frac{2}{u_k^2+\frac{1}{4}} = 4\,\left[S_1(N)-S_{-1}(N)\right].
\ee
As an easy check, we see that \refeq{h2} is in agreement with \refeq{cusp}. Also, we notice the following
remarkable {\em shift symmetry} (more on this in App.~(\ref{app:shift}))
\be
\gamma_2^{\rm ABA}(2\,n+1) = \gamma_2^{\rm ABA}(2\,n+2),\qquad n\in\mathbb{N}.
\ee

\subsection{The four-loop ABA result}

The four-loop ABA results can in principle be obtained analytically from the next-to-leading order Baxter equation
which is presented in App.~(\ref{app:NLOBaxter}). It can be much more easily be determined assuming maximal and uniform  transcendentality 
of the participating harmonic sums and matching this Ansatz to the perturbative solution of the Bethe equations. We obtain the nice
result
\ba
\label{eq:h4}
\gamma_4^{\rm ABA}(N) &=& -16 (S_{-3}-S_3+S_{-2,-1}-S_{-2,1}+S_{-1,-2}-S_{-1,2}-S_{1,-2}+S_{1,2}-S_{2,-1}+S_{2,1}+\nonumber\\
&& + S_{-1,-1,-1}-S_{-1,-1,1}-S_{1,-1,-1}+S_{1,-1,1}).
\ea
Again, as a check, we agree with \refeq{cusp}. Also, and remarkably, \refeq{h4} still enjoys the {\em shift symmetry}
\be
\gamma_4^{\rm ABA}(2\,n+1) = \gamma_4^{\rm ABA}(2\,n+2),\qquad n\in\mathbb{N}.
\ee
Finally, as a further feature, we remark the following simple coefficient pattern
\be
\gamma_4^{\rm ABA} = \sum_{\mathbf{a}\in{\cal A}}c_\mathbf{a}\,S_\mathbf{a},
\ee
where $\mathbf{a}$ is a multi-index with \tran = 3, ${\cal A}$ is a proper subset of all multi-indices with \tran = 3, 
and 
\be
c_{a_1, \dots, a_\ell} = 16\,\prod_{i=1}^\ell \sigma_{a_i},
\ee
where
\be
\sigma_a = \left\{\begin{array}{ll}
+1 & \qquad a\ {\rm odd\ positive\ or\ even\ negative}\\
-1 & \qquad {\rm otherwise}
\end{array}\right.
\ee
Notice that this pattern works also for the two-loop result with trivial modifications.

\subsection{The four-loop wrapping contribution}

The full anomalous dimension of twist-1 operators receives a wrapping contribution at four loops
\be
\gamma_4(N) = \gamma_4^{\rm ABA}(N) + \gamma_4^{\rm wrap}(N).
\ee
We have worked out the wrapping contribution according to the proposal in\cite{Gromov:2009tv}~\footnote{M. B. thanks 
Nikolay Gromov and Pedro Vieira for major help in deriving the detailed form of ${\cal W}(N)$.}. In that framework one obtains
\be
\gamma_4^{wrap} = \gamma_2(N)\cdot{\cal W}(N),
\ee
with 
\ba
{\cal W}(N) &=& \sum_{Q=1}^\infty\int_{-\infty}^\infty dq\,{\cal W}(q, Q, N), \\
{\cal W}(q, Q, N) &=& -\frac{1}{2\,\pi}\,\frac{4}{q^2+Q^2}\,{\cal S}(q, Q, N)\,{\cal M}(q, Q, N), \nonumber
\ea
and ($Q_N$ is the leading order Baxter function)
\ba
{\cal S}(q, Q, N) &=& (-1)^Q\,\frac{Q_N\left(\frac{q-i(Q-1)}{2}\right)}
{Q_N\left(\frac{q-i(Q+1)}{2}\right) Q_N\left(\frac{q+i(Q-1)}{2}\right) Q_N\left(\frac{q+i(Q+1)}{2}\right)}, \\
{\cal M}(q, Q, N) &=& 2\,\sum_{j=0}^{Q-1}\left[\frac{Q_N\left(\frac{q-i(Q-1)+2\,i\,j}{2}\right)}
{Q_N\left(\frac{q-i(Q-1)}{2}\right)}\right]^2\,\left[\frac{1}{2\,j-i\,q-Q}-\,\frac{1}{2\,(j+1)-i\,q-Q}\right].\nonumber
\ea
This formula takes into account the different $SU(2|2)$ structure of the $S$-matrix as compared with ${\cal N}=4$.
Under summation over $Q$, the integral can be evaluated in terms of the kinematical residue 
\be
{\cal W}(N) = 2\,\pi\,i\,\sum_{Q=1}^\infty \mathop{\mbox{Res}}_{q=i\,Q} {\cal W}(q, Q, N).
\ee
This sum can be computed for each $N$ and it takes the form 
\be
{\cal W}(N) = -2\,\zeta_2 + r_N,
\ee
where $r_N$ is a rational number. Unfortunately, we have been unable to find a closed formula for $r_N$. However, we can show that at large $N$ 
the leading term in ${\cal W}$ is 
\be
\label{eq:leading}
{\cal W}(N) = -\frac{2\,\log\,2}{N} + \mbox{subleading}.
\ee
In particular, this proves that the cusp anomaly is not modified by the wrapping which goes like $(\log\,N)/N$ at large $N$. 
To show this important fact, we first observe that 
\be
R_Q(N) = 2\,\pi\,i\, \mathop{\mbox{Res}}_{q=i\,Q} {\cal W}(q, Q, N) = \frac{A_Q(N)}{B_Q(N)^2}, 
\ee
where $A_Q(N)$ and $B_Q(N)$ are polynomials with 
\be
\mbox{deg}\,A_Q(N) = 4\,Q-3,\qquad\mbox{deg}\,B_Q(N) = 2\,Q-1.
\ee
This statement can be proved starting from the Baxter equation and the residue formula.
The first cases (for even $N$) are 
\ba
R_1(N) &=&  -\frac{4 (N+1)}{(2 N+1)^2}, \nonumber\\
R_2(N) &=& -\frac{8 N^5+20 N^4+26 N^3+15 N^2+2 N-1}{(2 N+1)^2 \left(2 N^2+2 N+1\right)^2},  \\
R_3(N) &=& \frac{4 \left(8 N^9+36 N^8+112 N^7+224 N^6+342 N^5+367 N^4+268 N^3+114 N^2+17 N-3\right)}{3 (2 N+1)^2 \left(2 N^2+2 N+1\right)^2 \left(2 N^2+2 N+3\right)^2}
\nonumber
\ea
The large $N$ expansion of these rational functions is 
\ba
R_1(N) &=& -\frac{1}{N}+\frac{1}{4} \,\frac{1}{N^3}-\frac{1}{4} \,\frac{1}{N^4}+\cdots, \nonumber \\
R_2(N) &=& -\frac{1}{2 N}+\frac{1}{4} \,\frac{1}{N^2}-\frac{1}{4} \,\frac{1}{N^3}+\frac{1}{2}
   \,\frac{1}{N^4}+\cdots,\nonumber \\
R_3(N) &=& \frac{1}{6 N}-\frac{1}{12} \,\frac{1}{N^2}+\frac{3}{8} \,\frac{1}{N^3}-\frac{25}{48}
   \,\frac{1}{N^4}+\cdots,\\
R_4(N) &=& -\frac{1}{12 N}+\frac{1}{24} \,\frac{1}{N^2}-\frac{5}{24} \,\frac{1}{N^3}+\frac{7}{24}
   \,\frac{1}{N^4}+\cdots,\nonumber \\
R_5(N) &=& \frac{1}{20 N}-\frac{1}{40} \,\frac{1}{N^2}+\frac{7}{48} \,\frac{1}{N^3}-\frac{33}{160}
   \,\frac{1}{N^4}+\cdots,\nonumber \\
R_6(N) &=& -\frac{1}{30 N}+\frac{1}{60} \,\frac{1}{N^2}-\frac{9}{80} \,\frac{1}{N^3}+\frac{77}{480}
   \,\frac{1}{N^4}+\cdots.\nonumber
\ea
The leading term can be written 
\be
R_Q(N) = \frac{c_Q}{N} + {\cal O}(N^{-2}),
\ee
and we have checked that up to large $Q$ one has 
\be
c_1 = -1,\qquad c_{Q>1} = -\frac{(-1)^Q}{Q\,(Q-1)}.
\ee
This means that we can analytically compute the sum over $Q$ 
\be
\sum_{Q=1}^\infty c_Q = -1-\sum_{Q=2}^\infty \frac{(-1)^Q}{Q\,(Q-1)} = -2\,\log\,2,
\ee
and we get \refeq{leading}.

\subsection{The six-loop ABA result}

With a certain effort, and using again maximal transcendentality, we have obtained the following six-loop formula
{\small 
\ba
\label{eq:h6}
\gamma_6^{\rm ABA}(N) &=& 
-16 (8 S_{-5}-8 S_5+14 S_{-4,-1}-14 S_{-4,1}+24 S_{-3,-2}-24 S_{-3,2}+22 S_{-2,-3}-22 S_{-2,3}+\nonumber\\
&& +16 S_{-1,-4}-16 S_{-1,4}-12 S_{1,-4}+12 S_{1,4}-26 S_{2,-3}+26
   S_{2,3}-24 S_{3,-2}+24 S_{3,2}+\nonumber\\
&& -14 S_{4,-1}+14 S_{4,1}+24 S_{-3,-1,-1}-24 S_{-3,-1,1}-12 S_{-3,1,-1}+12 S_{-3,1,1}+21 S_{-2,-2,-1}+\nonumber\\
&& -21 S_{-2,-2,1}+33
   S_{-2,-1,-2}-33 S_{-2,-1,2}-9 S_{-2,1,-2}+9 S_{-2,1,2}-21 S_{-2,2,-1}+21 S_{-2,2,1}+\nonumber\\
&& +22 S_{-1,-3,-1}-22 S_{-1,-3,1}+30 S_{-1,-2,-2}-30 S_{-1,-2,2}+32
   S_{-1,-1,-3}-32 S_{-1,-1,3}+\nonumber\\
&& -12 S_{-1,1,-3}+12 S_{-1,1,3}-30 S_{-1,2,-2}+30 S_{-1,2,2}-22 S_{-1,3,-1}+22 S_{-1,3,1}-14 S_{1,-3,-1}+\nonumber\\
&& +14 S_{1,-3,1}-18
   S_{1,-2,-2}+18 S_{1,-2,2}-20 S_{1,-1,-3}+20 S_{1,-1,3}+8 S_{1,1,-3}-8 S_{1,1,3}+18 S_{1,2,-2}+\nonumber\\
&& -18 S_{1,2,2}+14 S_{1,3,-1}-14 S_{1,3,1}-27 S_{2,-2,-1}+27
   S_{2,-2,1}-39 S_{2,-1,-2}+39 S_{2,-1,2}+15 S_{2,1,-2}+\nonumber\\
&& -15 S_{2,1,2}+27 S_{2,2,-1}-27 S_{2,2,1}-24 S_{3,-1,-1}+24 S_{3,-1,1}+12 S_{3,1,-1}-12 S_{3,1,1}+\nonumber\\
&& + 34
   S_{-2,-1,-1,-1}-34 S_{-2,-1,-1,1}-16 S_{-2,-1,1,-1}+16 S_{-2,-1,1,1}-10 S_{-2,1,-1,-1}+\nonumber\\
&& + 10 S_{-2,1,-1,1}+32 S_{-1,-2,-1,-1}-32 S_{-1,-2,-1,1}-16
   S_{-1,-2,1,-1}+16 S_{-1,-2,1,1}+36 S_{-1,-1,-2,-1}+\nonumber\\
&& -36 S_{-1,-1,-2,1}+50 S_{-1,-1,-1,-2}-50 S_{-1,-1,-1,2}-22 S_{-1,-1,1,-2}+22 S_{-1,-1,1,2}-36
   S_{-1,-1,2,-1}+\nonumber\\
&& +36 S_{-1,-1,2,1}-12 S_{-1,1,-2,-1}+12 S_{-1,1,-2,1}-22 S_{-1,1,-1,-2}+22 S_{-1,1,-1,2}+2 S_{-1,1,1,-2}+\nonumber\\
&& -2 S_{-1,1,1,2}+12 S_{-1,1,2,-1}-12
   S_{-1,1,2,1}-32 S_{-1,2,-1,-1}+32 S_{-1,2,-1,1}+16 S_{-1,2,1,-1}-16 S_{-1,2,1,1}+\nonumber\\
&& -18 S_{1,-2,-1,-1}+18 S_{1,-2,-1,1}+6 S_{1,-2,1,-1}-6 S_{1,-2,1,1}-18
   S_{1,-1,-2,-1}+18 S_{1,-1,-2,1}+\nonumber\\
&& -28 S_{1,-1,-1,-2}+28 S_{1,-1,-1,2}+8 S_{1,-1,1,-2}-8 S_{1,-1,1,2}+18 S_{1,-1,2,-1}-18 S_{1,-1,2,1}+\nonumber\\
&& +6 S_{1,1,-2,-1}-6
   S_{1,1,-2,1}+12 S_{1,1,-1,-2}-12 S_{1,1,-1,2}-6 S_{1,1,2,-1}+6 S_{1,1,2,1}+18 S_{1,2,-1,-1}+\nonumber\\
&& -18 S_{1,2,-1,1}-6 S_{1,2,1,-1}+6 S_{1,2,1,1}-40 S_{2,-1,-1,-1}+40
   S_{2,-1,-1,1}+22 S_{2,-1,1,-1}-22 S_{2,-1,1,1}+\nonumber\\
&& +16 S_{2,1,-1,-1}-16 S_{2,1,-1,1}-6 S_{2,1,1,-1}+6 S_{2,1,1,1}+52 S_{-1,-1,-1,-1,-1}-52 S_{-1,-1,-1,-1,1}+\nonumber\\
&& -32
   S_{-1,-1,-1,1,-1}+32 S_{-1,-1,-1,1,1}-24 S_{-1,-1,1,-1,-1}+24 S_{-1,-1,1,-1,1}+12 S_{-1,-1,1,1,-1}+\nonumber\\
&& -12 S_{-1,-1,1,1,1}-24 S_{-1,1,-1,-1,-1}+24
   S_{-1,1,-1,-1,1}+12 S_{-1,1,-1,1,-1}-12 S_{-1,1,-1,1,1}+\nonumber\\
&& + 4 S_{-1,1,1,-1,-1}-4 S_{-1,1,1,-1,1}-28 S_{1,-1,-1,-1,-1}+28 S_{1,-1,-1,-1,1}+12 S_{1,-1,-1,1,-1}+\nonumber\\
&& -12
   S_{1,-1,-1,1,1}+8 S_{1,-1,1,-1,-1}-8 S_{1,-1,1,-1,1}+12 S_{1,1,-1,-1,-1}-12 S_{1,1,-1,-1,1}+\nonumber\\
&& -4 S_{1,1,-1,1,-1}+4 S_{1,1,-1,1,1}).
\ea
}
This formula is in agreement with \refeq{cusp}, but the {\em shift symmetry} is broken. We do not know whether it must be an all-order 
property of the anomalous dimension. To any extent, the six loop result is affected by next-to-leading wrapping contributions which are 
not known even in ${\cal N}=4$. Therefore, we shall avoid any analysis of this result which, in our opinion, deserves a better treatment of wrapping effects.

\section{Twist-2}
\subsection{The two-loop problem}

The two-loop problem is not discussed in~\cite{Zwiebel:2009vb} and our results are simple, but new. 
The Bethe equations are ({\bf even} $N$) 
\be
\left(\frac{u_k+\frac{i}{2}}{u_k-\frac{i}{2}}\right)^2 = -\prod_{j\neq k}^N\frac{u_k-u_j-i}{u_k-u_j+i},\qquad k,j = 1, \dots, N.
\ee
They turn out to be equivalent to the following leading order Baxter equation 
\be
\left(u+\frac{i}{2}\right)^2\,Q_N(u+i)-\left(u-\frac{i}{2}\right)^2\,Q_N(u-i) = i\,(2\,N+2)\,u\,Q_N(u),
\ee
for the Baxter polynomial associated to the ground state $Q_N(u) = {\cal N}\,\prod_{k=1}^N (u-u_k)$.
The physical solution is easily found to be 
\be
Q_N(u) ={}_3F_2\left(\left.\begin{array}{c}
-\frac{N}{2}, \quad   i\,u+\frac{1}{2}, \quad -i\,u+\frac{1}{2} \\
1,\quad 1
\end{array}
\right| 1\right).
\ee
Therefore, the anomalous dimension can be computed exactly and reads
\be
\label{eq:h22}
\gamma_2^{\rm ABA}(N) = \sum_k\frac{2}{u_k^2+\frac{1}{4}} = 4\,\left[S_1(N)+S_{-1}(N)\right].
\ee
\refeq{h22} is in agreement with \refeq{cusp}. As in the twist-1 case, there is a {\em shift symmetry} since 
$\gamma_2^{\rm ABA}$ enjoys the exact property
\be
\gamma_2^{\rm ABA}(2\,n+1) = \gamma_2^{\rm ABA}(2\,n),\qquad n\in\mathbb{N}.
\ee

\subsection{The four-loop ABA result}

After some calculation, we obtain 
\ba
\label{eq:h24}
\gamma_4^{ABA}(N) &=& 16 S_{-3}+16 S_3-8 S_{-2,-1}-8 S_{-2,1}-16 S_{-1,-2} + \nonumber\\
&& -16 S_{-1,2}-16 S_{1,-2}-16 S_{1,2}-8 S_{2,-1}-8 S_{2,1}.
\ea
\refeq{h24} is proved in the Appendix by means of the NLO Baxter equation. It is in agreement with \refeq{cusp} and has again the exact property
\be
\gamma_4^{\rm ABA}(2\,n+1) = \gamma_4^{\rm ABA}(2\,n),\qquad n\in\mathbb{N}.
\ee

\subsection{The six-loop ABA result}

At six-loop, we obtain 
{\small 
\ba
\label{eq:h26}
\gamma_6^{\rm ABA}(N) &=& 
128 \left(S_{-5}+S_5\right)-192 \left(S_{-1,-4}+S_{1,-4}\right)-192 \left(S_{-1,4}+S_{1,4}\right)-256 \left(S_{-2,-3}+S_{2,-3}\right) \nonumber\\
&& -256
   \left(S_{-2,3}+S_{2,3}\right)-160 \left(S_{-3,-2}+S_{3,-2}\right)-160 \left(S_{-3,2}+S_{3,2}\right)-128 \left(S_{-4,-1}+S_{4,-1}\right) \nonumber\\
&& -128
   \left(S_{-4,1}+S_{4,1}\right)+96 \left(S_{-1,-3,-1}+S_{1,-3,-1}\right)+96 \left(S_{-1,-3,1}+S_{1,-3,1}\right) \nonumber\\
&& +96 \left(S_{-1,-2,-2}+S_{1,-2,-2}\right)+96
   \left(S_{-1,-2,2}+S_{1,-2,2}\right)+128 \left(S_{-1,-1,-3}+S_{1,-1,-3}\right)\nonumber\\
&& +128 \left(S_{-1,-1,3}+S_{1,-1,3}\right)+128
   \left(S_{-1,1,-3}+S_{1,1,-3}\right)+128 \left(S_{-1,1,3}+S_{1,1,3}\right)\nonumber\\
&& +96 \left(S_{-1,2,-2}+S_{1,2,-2}\right)+96 \left(S_{-1,2,2}+S_{1,2,2}\right)+96
   \left(S_{-1,3,-1}+S_{1,3,-1}\right)+96 \left(S_{-1,3,1}+S_{1,3,1}\right)\nonumber\\
&& +96 \left(S_{-2,-2,-1}+S_{2,-2,-1}\right)+96 \left(S_{-2,-2,1}+S_{2,-2,1}\right)+80
   \left(S_{-2,-1,-2}+S_{2,-1,-2}\right)\nonumber\\
&& +80 \left(S_{-2,-1,2}+S_{2,-1,2}\right)+80 \left(S_{-2,1,-2}+S_{2,1,-2}\right)+80 \left(S_{-2,1,2}+S_{2,1,2}\right)\nonumber\\
&& +96
   \left(S_{-2,2,-1}+S_{2,2,-1}\right)+96 \left(S_{-2,2,1}+S_{2,2,1}\right)+32 \left(S_{-3,-1,-1}+S_{3,-1,-1}\right)\nonumber\\
&& +32 \left(S_{-3,-1,1}+S_{3,-1,1}\right)+32
   \left(S_{-3,1,-1}+S_{3,1,-1}\right)+32 \left(S_{-3,1,1}+S_{3,1,1}\right)\nonumber\\
&& -32 \left(S_{-1,-1,-2,-1}+S_{1,-1,-2,-1}\right)-32
   \left(S_{-1,-1,-2,1}+S_{1,-1,-2,1}\right)-32 \left(S_{-1,-1,2,-1}+S_{1,-1,2,-1}\right)\nonumber\\
&& -32 \left(S_{-1,-1,2,1}+S_{1,-1,2,1}\right)-32
   \left(S_{-1,1,-2,-1}+S_{1,1,-2,-1}\right)-32 \left(S_{-1,1,-2,1}+S_{1,1,-2,1}\right)\nonumber\\
&& -32 \left(S_{-1,1,2,-1}+S_{1,1,2,-1}\right)-32
   \left(S_{-1,1,2,1}+S_{1,1,2,1}\right).
\ea
}
\refeq{h26} is in agreement with \refeq{cusp}. Remarkably, {\em shift symmetry} is not broken. Wrapping effects are expected to show up 
at this order.

\subsection{A convenient reshuffling}

Inspired by what happens in the case of ${\cal N}=4$ twist-3 fields, we can rewrite the ABA results 
in terms of harmonic sums with positive indices and argument $M\equiv N/2$~\footnote{We are indebted to T. Lukowski for 
discussions on this point.}. We find the simple expressions
\ba
S_\mathbf{a} &\equiv& S_\mathbf{a}(M), \\
\gamma_2^{\rm ABA} &=& 4\,S_1, \\
\gamma_4^{\rm ABA} &=& 4\, S_3-8\, S_{1,2}-4\, S_{2,1}, \\
\gamma_6^{\rm ABA} &=& 8 \,S_5-24 \,S_{1,4}-32 \,S_{2,3}-20 \,S_{3,2}-16 \,S_{4,1}+32 \,S_{1,1,3}+24 \,S_{1,2,2}+\nonumber\\
&& + 24 \,S_{1,3,1}+20 \,S_{2,1,2}+24 \,S_{2,2,1}+8 \,S_{3,1,1}-16 \,S_{1,1,2,1}.
\ea

\section{Reciprocity and LBK wisdom: Large $N$ analysis}

We can now come back to our main aim which is the analysis of possible QCD-inspired properties showing up in $\gamma_{2n}^{\rm ABA}(N)$~\footnote{We
cannot fully analyze the dressing contribution since a closed formula with the complete dependence on $N$ is not available, yet.}.
Mimicking the ${\cal N}=4$ case, we shall work out the expansion of the anomalous dimensions at large $N$ and look for peculiar properties~\footnote{Technically, the 
details of the expansion are fully discussed in~\cite{Beccaria:2009vt,Albino:2009ci}.}. For generic $N$ one can write 
\be
\gamma(N) = \alpha(N) + (-1)^N\,\beta(N),
\ee
where $\alpha$ and $\beta$ have a smooth expansion in $1/N$ with possible logarithmic enhancements. We shall consider the even $N$ case
which turns out to be the most interesting. The general form of the large $N$ expansion is expected to be 
\be
\label{eq:gexp}
\gamma(N) = f_{CS}(h)\,\log\,N + \sum_{a=1}^\infty\frac{1}{N^a}\sum_{b=0}^a\,g_{a,b}(h)\,\log^b N.
\ee
We have already checked that the leading cusp logarithm is in agreement with property (a) discussed in the Introduction. So, as expected, logarithmic scaling 
works. Concerning properties (b)-(c), they are conveniently expressed in terms of the function $P$ defined order by order in $h$ by the functional
relation~\cite{Dokshitzer:2005bf,Dokshitzer:2006nm,Basso:2006nk}
\be
\gamma(N) = P\left(N + \frac{1}{2}\,\gamma(N)\right).
\ee
The large $N$ expansion of $P$ is similar to that of $\gamma$ and reads
\be
\label{eq:pexp}
P(N) = f_{CS}(h)\,\log\,N + \sum_{a=1}^\infty\frac{1}{N^a}\sum_{b=0}^a\,p_{a,b}(h)\,\log^b N.
\ee
The Gribov-Lipatov reciprocity and LBK cancellations can be concisely expressed as follows.
\begin{enumerate}
\item[-] {\em Gribov-Lipatov reciprocity.} There is a constant $\kappa$ such that the large $N$ expansion of $P(N)$ runs in 
integer inverse powers of $J^2 = N\,(N+\kappa)$.

\item[-] {\em Low-Burnett-Kroll cancellations.} Some (maximal) logarithms are missing in \refeq{pexp}. This implies that there are inheritance
relations among the logarithms of \refeq{gexp}.
\end{enumerate}
We recall once again that these seemingly technical conditions have a clear physical origin in the QCD context and are widely checked in ${\cal N}=4$.
It remains to look for their manifestation in ABJM, at least at the level of the multi-loop asymptotic anomalous dimensions.

\subsection{Twist-1}

We define $\overline{n} = N\,e^{\gamma_E}$ and consider even $N$. The expansion of the two loop anomalous dimensions is 
\ba
\gamma_2^{\rm ABA} &=& 4 \log (2\, \overline{n})+\frac{2}{3\,n^2}-\frac{7}{15\,n^4}+\frac{62}{63\,n^6}-\frac{127}{30\,n^8}+\cdots.
\ea
At four loops, we find instead
\ba
\gamma_4^{\rm ABA} &=& \left(-\frac{4}{3} \pi ^2 \log (2\,\overline{n})-12 \zeta_3\right)+\frac{8 \log (2\,\overline{n})}{n}+\left(2 \log (2\,\overline{n})-\frac{2 \pi ^2}{9}+2\right)
   \frac{1}{n^2}+\\
&& + \left(\frac{4}{3}-\frac{8}{3} \log (2\,\overline{n})\right) \frac{1}{n^3}+\left(-\frac{5}{2} \log (2 \, \overline{n})
+\frac{7 \pi ^2}{45}+\frac{1}{12}\right) \frac{1}{n^4}+\nonumber \\
&& + \left(\frac{56}{15} \log (2\,\overline{n})-\frac{62}{45}\right)
   \frac{1}{n^5}+\left(7 \log (2\,\overline{n})-\frac{62 \pi ^2}{189}-\frac{269}{60}\right)
   \frac{1}{n^6}+\nonumber \\
&& + \left(\frac{914}{315}-\frac{248}{21} \log (2\,\overline{n})\right) \frac{1}{n^7}+\left(-\frac{285}{8} \log (2
   \,\overline{n})+\frac{127 \pi ^2}{90}+\frac{76613}{2016}\right) \frac{1}{n^8}+\cdots\nonumber 
\ea
As we remarked, for twist-1, we do not discuss the six-loop result which is heavily affected by the wrapping corrections. The two loop result is parity invariant under
the transformation ($\kappa=0$ in the Gribov-Lipatov reciprocity) 
\ba
n &\to& -n, \\
\log n &=& \frac{1}{2}\,\log (n^2) \to \log n.
\ea
This is not a symmetry of the four loop result. Nevertheless, we can look at the four loop $P$ function 
($P = \sum_{n=1}^\infty P_{2n}\,h^{2n}$)
\be
P_4 = \gamma_4 -\frac{1}{2}\,\gamma_2\,\gamma_2'.
\ee
After a brief calculation, we find 
\ba
P_4 &=& 2 \pi ^2 S_{-1}-2 \pi ^2 S_1-8 S_{-2,-1}+8 S_{-2,1}-8 S_{-1,-2}+8 S_{-1,2}+8 S_{1,-2}-8 S_{1,2}+8 S_{2,-1}+\nonumber\\
&& -8 S_{2,1}-16 S_{-1,-1,-1}+16 S_{-1,-1,1}+16
   S_{1,-1,-1}-16 S_{1,-1,1},
\ea
and its expansion is 
\ba
P_4 &=& 
\left(-\frac{4}{3} \pi ^2 \log (2\,\overline{n})-12 \zeta_3\right)+\left(2 \log (2\,\overline{n})-\frac{2 \pi ^2}{9}+2\right)
   \frac{1}{n^2}+\\
&& + \left(-\frac{5}{2} \log (2\,\overline{n})+\frac{7 \pi ^2}{45}+\frac{1}{12}\right) \frac{1}{n^4}+\left(7 \log (2\,\overline{n})
-\frac{62 \pi ^2}{189}-\frac{269}{60}\right) \frac{1}{n^6}+\nonumber\\
&& + \left(-\frac{285}{8} \log (2\,\overline{n})+\frac{127 \pi
   ^2}{90}+\frac{76613}{2016}\right) \frac{1}{n^8}+\cdots\nonumber
\ea
We see that $P_4$ is indeed parity invariant ! This structure implies that all terms in $\gamma_4$ odd under $n\to -n$ 
are precisely {\em inherited} from the two-loop anomalous dimension 
\be
\label{eq:restwist1}
\gamma_4^{\rm odd} = \frac{1}{2}\,\gamma_2\,\gamma_2'.
\ee
We can summarize the result \refeq{restwist1} by saying that {\em the twist-1 ABA anomalous dimension is reciprocity respecting under $n\to -n$ up to four loops}.
Instead, no LBK cancellation is observed. The logarithmic enhancement which are observed in $\gamma_4$ are the same as in $P_4$. This means that the single logarithms
appearing in $\gamma_4$ are not related to the lowest order $\gamma_2$.

\subsection{Twist-2}

For twist-2, we use the variable $n=N/2$ and we find,  with $\overline n = n\,e^{\gamma_E}$, the expansions at 2 and 4 loops 
\ba
\gamma_2^{\rm ABA} &=& 4 \log {\overline n}+\frac{2}{n}-\frac{1}{3\,n^2}+\frac{1}{30\,n^4}-\frac{1}{63\,n^6}+\frac{1}{60\,n^8}+\cdots, \\
\gamma_4^{\rm ABA} &=& \left(4 \zeta_3-\frac{4}{3} \pi ^2 \log {\overline n}\right)+\left(4 \log {\overline n}-\frac{2 \pi ^2}{3}-4\right) \frac{1}{n}+\left(-2 \log {\overline n}+\frac{\pi
   ^2}{9}+5\right) \frac{1}{n^2}+\nonumber\\
&& + \left(\frac{2 \log {\overline n}}{3}-\frac{29}{9}\right) \frac{1}{n^3}+\left(\frac{4}{3}-\frac{\pi
   ^2}{90}\right) \frac{1}{n^4}+\left(-\frac{2 \log {\overline n}}{15}-\frac{19}{225}\right)
   \frac{1}{n^5}+\nonumber\\
&& + \left(-\frac{19}{60}+\frac{\pi ^2}{189}\right) \frac{1}{n^6}+\left(\frac{2 \log
   {\overline n}}{21}+\frac{41}{588}\right) \frac{1}{n^7}+\left(\frac{17}{63}-\frac{\pi ^2}{180}\right)
   \frac{1}{n^8}+\cdots,
\ea
as well as the six-loop result
\ba
\gamma_6^{\rm ABA} &=& \left(\frac{44}{45} \pi ^4 \log {\overline n}-88 \zeta_5\right)+\frac{-\frac{10}{3} \pi ^2 \log {\overline n}+16 \log {\overline n}+4 \zeta_3+\frac{22 \pi
   ^4}{45}+\frac{2 \pi ^2}{3}+48}{n}+\nonumber\\
&& + \left(-2 \log ^2{\overline n}+\frac{5}{3} \pi ^2 \log {\overline n}+2 \log {\overline n}-2 \zeta_3-\frac{11 \pi
   ^4}{135}-\frac{5 \pi ^2}{2}-20\right) \frac{1}{n^2}+\nonumber\\
&& + \left(2 \log ^2{\overline n}-\frac{5}{9} \pi ^2 \log {\overline n}-\frac{292 \log
   {\overline n}}{27}+\frac{2 \zeta_3}{3}+\frac{95 \pi ^2}{54}+\frac{446}{27}\right) \frac{1}{n^3}+\nonumber\\
&& + \left(-\log ^2{\overline n}+\frac{193 \log
   {\overline n}}{18}+\frac{11 \pi ^4}{1350}-\frac{13 \pi ^2}{18}-\frac{416}{27}\right) \frac{1}{n^4}+\nonumber\\
&& + \left(\frac{1}{9} \pi ^2 \log
   {\overline n}-\frac{19007 \log {\overline n}}{3375}-\frac{2 \zeta_3}{15}+\frac{13 \pi ^2}{450}+\frac{2365123}{202500}\right)
   \frac{1}{n^5}+\nonumber\\
&& + \left(\frac{\log ^2{\overline n}}{3}+\frac{13 \log {\overline n}}{150}-\frac{11 \pi ^4}{2835}+\frac{59 \pi
   ^2}{360}-\frac{19603}{3375}\right) \frac{1}{n^6}+\nonumber\\ 
 && + \left(-\frac{5}{63} \pi ^2 \log {\overline n}+\frac{1006399 \log {\overline n}}{463050}+\frac{2
   \zeta_3}{21}-\frac{643 \pi ^2}{52920}+\frac{21409879}{64827000}\right) \frac{1}{n^7}+\nonumber\\
&& + \left(-\frac{1}{3} \log ^2{\overline n}-\frac{107 \log
   ({\overline n} )}{588}+\frac{11 \pi ^4}{2700}-\frac{26 \pi ^2}{189}+\frac{917411}{463050}\right)
   \frac{1}{n^8}+\cdots.
\ea
Again, possible structures are best investigated by looking at the $P$ functions.
Using $M=N/2$ as argument and inverting the relation 
\be
\gamma^{\rm ABA}(M) = P\left(M + \frac{1}{4}\,\gamma^{\rm ABA}(M)\right),
\ee
we get the expressions 
\ba
P_2 &=& \gamma_2, \\
P_4 &=& \gamma_4-\frac{1}{4}\, \gamma_2\,\gamma_2', \\
P_6 &=& \gamma_6-\frac{1}{4}\,\gamma_4\,\gamma_2'+\frac{1}{16}\,\gamma_2\,(\gamma_2')^2-\frac{1}{4}\,\gamma_2\,\gamma_4'+\frac{1}{32}\,\gamma_2^2\,\gamma_2''
\ea
Expanding at large $n$, we find 
\ba
P_2 &=& \gamma_2 = 4 \log {\overline n}+\frac{2}{n}-\frac{1}{3\,n^2}+\frac{1}{30\,n^4}-\frac{1}{63\,n^6}+\frac{1}{60\,n^8}+\cdots, \\
P_4 &=& \left(4 \zeta_3-\frac{4}{3} \pi ^2 \log \overline{n}\right)-\left(4+\frac{2 \pi ^2}{3}\right)\,\frac{1}{n}+\left(3+\frac{\pi ^2}{9}\right)
   \frac{1}{n^2}-\frac{17}{9} \frac{1}{n^3}+\left(\frac{5}{6}-\frac{\pi ^2}{90}\right) \frac{1}{n^4} + \nonumber \\
&& -\frac{14}{225}
   \frac{1}{n^5}+\left(-\frac{7}{30}+\frac{\pi ^2}{189}\right) \frac{1}{n^6}+\frac{152}{2205}\frac{1}{n^7}+\left(\frac{3}{14}-\frac{\pi ^2}{180}\right) \frac{1}{n^8}+\cdots, \nonumber\\
P_6 &=& \left(\frac{44 \pi ^4}{45} \log\overline{n} -88 \zeta_5\right)+\left(\left(16-\frac{2 \pi ^2}{3}\right) \log\overline{n}+\frac{22 \pi ^4}{45}+\frac{2 \pi
   ^2}{3}+48\right)\frac{1}{n}+\nonumber\\
&& \left(\left(-6+\frac{\pi ^2}{3}\right) \log\overline{n}-\frac{11 \pi ^4}{135}-\frac{7 \pi ^2}{6}-16\right)
   \frac{1}{n^2}+\nonumber\\
&& \left(\left(\frac{32}{27}-\frac{\pi ^2}{9}\right) \log\overline{n}+\frac{47 \pi ^2}{54}+\frac{203}{27}\right)
   \frac{1}{n^3}+\nonumber\\
&& \left(\frac{7}{18} \log\overline{n}+\frac{11 \pi ^4}{1350}-\frac{7 \pi ^2}{18}-\frac{140}{27}\right)
   \frac{1}{n^4}+\nonumber\\
&& \left(\left(-\frac{1007}{3375}+\frac{\pi ^2}{45}\right) \log\overline{n}+\frac{19 \pi ^2}{1350}+\frac{790123}{202500}\right)
   \frac{1}{n^5}+\cdots
\ea
We did not find any simple parity invariance analogous to what is found in ${\cal N}=4$. Nevertheless, 
LBK cancellations are present. Indeed, the structure of the logarithmic expansion is peculiar. 
Apart from the cusp anomaly, $\gamma_2$ has no logarithms, $\gamma_4$ has simple logarithms, and $\gamma_6$ has squared logarithms. 
Instead, $P_2$ and $P_4$ are logarithm-free, whereas $P_6$ has only simple logarithms. This implies that the leading logarithms in the anomalous dimension 
are all inherited from the lowest order $\gamma$. In more details, one can check the remarkable relations 
\ba
\gamma_4^{\rm ABA}(n) &=& \log\,n\, \left[-\frac{4}{3}\,\pi^2 +\frac{d\gamma_2^{\rm ABA}(n)}{dn}\right] + {\cal O}(1), \\
\gamma_6^{\rm ABA}(n) &=& \frac{1}{2} \log^2 n\,\frac{d^2\gamma_2^{\rm ABA}(n)}{dn^2} + {\cal O}(\log\,n).
\ea
We can summarize this result by saying that {\em the twist-2 ABA anomalous dimension has leading order LBK inheritance up to six loops}.

\section{Conclusions}

The ABJM theory shares many similarities with ${\cal N}=4$ SYM at strong coupling where the dual $AdS_4\times \mathbb{CP}^3$ replaces the maximal 
background $\ads$. Instead, at weak coupling, the two theories have a quite different structure. Nevertheless, the all-loop conjectured Bethe equations
keep showing strong similarities. In particular, there is a $\mathfrak{sl}(2)$ sector in both cases with very closely related Bethe Ansatz equations.
In ${\cal N}=4$, this non-compact sector contains nice twist operators with a prominent role in the attempt to exchange ideas and conjectures with the
observed ${\cal N}=0$ physics.

It is natural and puzzling to ask whether QCD-inspired physical properties of ${\cal N}=4$ SYM twist operators are robust enough to 
carry over to the ABJM context. This can only be possible if the structural similarity between the two theories is enough powerful.
In this paper, we have focused on two non-trivial features of ${\cal N}=4$ SYM twist operators, Gribov-Lipatov reciprocity and Low-Burnett-Kroll cancellations.
We have shown that these properties show up in a much softer and broken way compared to ${\cal N}=4$. Nevertheless, various intriguing remnants of these 
physical properties are still found in ABJM.

Indeed, the multi-loop analysis of the (asymptotic) anomalous dimensions of twist-1 and 2 operators reveals a curious pattern. Twist-1 operators obey a four loop
parity invariance closely related to conventional Gribov-Lipatov reciprocity. Instead, twist-2 operators have no non-trivial parity invariance, but 
display a variety of LBK cancellations up to six loops.

Were it not for the ${\cal N}=4$ case, one could naively conclude that these features are accidental. We believe that this is not the case. Technically, 
they are due to the partial similarity of the ABJM and ${\cal N}=4$ $\mathfrak{sl}(2)$ sectors. Physically, it would be very interesting to 
look for arguments leading to Gribov-Lipatov reciprocity and Low-Burnett-Kroll wisdom in the case of the ${\cal N}=6$ superconformal Chern-Simons theory.

A final comment is deserved by wrapping corrections. We have applied the Kazakov-Gromov-Vieira formalism to the evaluation of the leading wrapping correction
to twist-1 operators. It would be very interesting to work out closed formulae for this correction as well as explicit diagrammatic checks. 
If the proposed (conjectured) correction turns out to be correct, then our analysis confirms that wrapping is subleading at large $N$ 
as in ${\cal N}=4$ SYM. 

\section*{Acknowledgments}
M.~B. thanks for very useful discussions N. Gromov, P. Vieira, and T. Lukowski on the wrapping corrections in ABJM, 
S. Zieme and A. Belitsky on the analytical solutions to the Baxter equation, and B. Zwiebel on shift symmetry.
We also thank F.~L.~Alday, D.~Bak, S.-J.~Rey, J.~Gomis, D.~Sorokin, L.~Wulff, and D.~Fioravanti for comments.

\appendix
\section{On shift symmetries}
\label{app:shift}

Let us consider a linear combination of nested (signed) harmonic sums
\be
S(n) = \sum_\mathbf{a} c_\mathbf{a}\,S_\mathbf{a}(n),\qquad \mathbf{a} = (a_1, a_2, \cdots).
\ee
Then we have
\begin{theorem}
$S(n)$ has the shift symmetry 
\be
S(2n+1) = S(2n+2),
\ee
if and only if 
\be
c_{a, \mathbf{b}} = -c_{-a, \mathbf{b}}.
\ee
\end{theorem}

\begin{theorem}
$S(n)$ has the shift symmetry 
\be
S(2n+1) = S(2n),
\ee
if and only if 
\be
c_{a, \mathbf{b}} = c_{-a, \mathbf{b}}.
\ee
\end{theorem}

\section{Next-to-leading Baxter equation}
\label{app:NLOBaxter}

We collect in this Appendix a few interesting results concerning the NLO Baxter equations for twist-1, 2 operators.
These are meant as an analytical support to maximal transcendentality conjectures and as a first step toward 
an extension to ABJM of the ${\cal N}=4$ results~\cite{BaxterLong}.

\subsection{Twist-1}

The lowest order Baxter equations is 
\be
\left(u+\frac{i}{2}\right)\,Q^{(0)}(u+i)-\left(u-\frac{i}{2}\right)\,Q^{(0)}(u-i) - i\,(2\,N+1)\,Q^{(0)}(u) = 0,
\ee
and the solution associated to the ground state is the degree $N$ polynomial
\be
Q^{(0)}(u) = {}_2F_1\left(\left.\begin{array}{c}
-N, \quad   i\,u+\frac{1}{2} \\
1
\end{array}
\right| 2 \right).
\ee
One can construct the NLO Baxter equation for $Q^{(1)}$ where $Q = Q^{(0)}+h^2\,Q^{(1)} + \dots$. It turns out to be 
\ba
\lefteqn{\left(u+\frac{i}{2}\right)\,Q^{(1)}(u+i)-\left(u-\frac{i}{2}\right)\,Q^{(1)}(u-i) - i\,(2\,N+1)\,Q^{(1)}(u) = } && \nonumber \\
&& \qquad\qquad a(N)\,\left[(u-i\,N)\,Q^{(0)}(u+i)-(u+i\,N)\,Q^{(0)}(u-i)\right] \nonumber \\
&&\qquad\qquad  +\frac{Q^{(0)}(u+i)-Q^{(0)}(u)}{u+i/2}-\frac{Q^{(0)}(u-i)-Q^{(0)}(u)}{u-i/2},
\ea
where
\be
a(N) = \frac{4}{2\,N+1}\left[S_1(N)-S_{-1}(N)\right],
\ee
and the solution is uniquely found in the space of polynomials of degree $N-2$.
If needed, the analytical NLO solution can be investigated by the methods of~\cite{Kotikov:2008pv}~\footnote{S. Zieme, private communication.}.
%~\footnote{However, we remark that non polynomial inhomogeneities
%(the last line) lead only to an implicit solution in terms of multiple sums over Stirling numbers which are not very much better than 
%a case by case solution. In particular, the analytic contribution to the anomalous dimension from such terms has not been derived yet.}.

\subsection{Twist-2}

The lowest order Baxter equations is 
\be
\left(u+\frac{i}{2}\right)^2\,Q^{(0)}(u+i)-\left(u-\frac{i}{2}\right)^2\,Q^{(0)}(u-i) - i\,(2\,N+2)\,u\,Q^{(0)}(u) = 0,
\ee
and the solution associated to the ground state is the degree $N\in 2\,\mathbb{N}$ polynomial
\be
Q^{(0)}(u) = {}_3F_2\left(\left.\begin{array}{c}
-\frac{N}{2}, \quad   \frac{1}{2}+i\,u, \quad \frac{1}{2}-i\,u \\
1,\quad 1
\end{array}
\right| 1\right).
\ee
Again, one can construct the NLO Baxter equation for $Q^{(1)}$ where $Q = Q^{(0)}+h^2\,Q^{(1)} + \dots$. It turns out to be 
\ba
\lefteqn{\left(u+\frac{i}{2}\right)^2\,Q^{(1)}(u+i)-\left(u-\frac{i}{2}\right)^2\,Q^{(1)}(u-i) - i\,(2\,N+2)\,u\,Q^{(1)}(u) = }\nonumber\\
&& -2\,i\,S_1(N/2)\,\left[\left(u+\frac{i}{2}\right)\,Q^{(0)}(u+i)+\left(u-\frac{i}{2}\right)\,Q^{(0)}(u-i)-2\,u\,Q^{(0)}(u)\right] \nonumber\\
&& +2\,\left[
Q^{(0)}(u+i)-Q^{(0)}(u-i)
\right].
\ea
The methods of~\cite{Kotikov:2008pv} can be easily adapted to this case due to the particularly simple form of the 
inhomogeneities. The final result for $Q^{(1)}$ can be expressed in deformed hypergeometric form 
\be
Q^{(1)}(u) = -3\,S_1(N/2)\,\left[G^{(1)}_1(u)+G^{(1)}_2(u)\right]-G^{(2)}_1(u)-G^{(2)}_2(u)-G^{(2)}_3(u),
\ee
where
\ba
G^{(1)}_1(u) &=& \left. \frac{\partial}{\partial\alpha}\,{}_3F_2\left(\left.\begin{array}{c}
-\frac{N}{2}, \quad   \frac{1}{2}+i\,u+\alpha, \quad \frac{1}{2}-i\,u \\
1,\quad 1
\end{array}
\right| 1\right)\right|_{\alpha=0}, \\
G^{(1)}_2(u) &=& \left. \frac{\partial}{\partial\alpha}\,{}_3F_2\left(\left.\begin{array}{c}
-\frac{N}{2}, \quad   \frac{1}{2}+i\,u, \quad \frac{1}{2}-i\,u+\alpha \\
1,\quad 1
\end{array}
\right| 1\right)\right|_{\alpha=0},
\ea
and
\ba
G^{(2)}_1(u) &=& \left. \frac{\partial^2}{\partial\alpha\partial\beta}\,{}_3F_2\left(\left.\begin{array}{c}
-\frac{N}{2}, \quad   \frac{1}{2}+i\,u+\alpha, \quad \frac{1}{2}-i\,u \\
1,\quad 1+\beta
\end{array}
\right| 1\right)\right|_{\alpha=\beta=0}, \\
G^{(2)}_2(u) &=& \left. \frac{\partial^2}{\partial\alpha\partial\beta}\,{}_3F_2\left(\left.\begin{array}{c}
-\frac{N}{2}, \quad   \frac{1}{2}+i\,u, \quad \frac{1}{2}-i\,u+\alpha \\
1,\quad 1+\beta
\end{array}
\right| 1\right)\right|_{\alpha=\beta=0}, \\
G^{(2)}_3(u) &=& \left. \frac{\partial^2}{\partial\alpha^2}\,{}_3F_2\left(\left.\begin{array}{c}
-\frac{N}{2}, \quad   \frac{1}{2}+i\,u+\alpha, \quad \frac{1}{2}-i\,u+\alpha \\
1,\quad 1
\end{array}
\right| 1\right)\right|_{\alpha=0}.
\ea
From this result we can easily recover the four loop formula.

\end{document}